\documentclass{aa}
\usepackage{graphicx}
\begin{document}
%\thesaurus{ }

\headnote{Research Note}
\title{Radio emission and the optical isophotal twist of radio-loud ellipticals}

\author
{Gopal-Krishna\inst{1}
	\and Amit\  R.\ Dhakulkar\inst{2}
	\and Paul J.\ Wiita\inst{3}
	\and Samir Dhurde\inst{2}
}
%\offprints{Gopal-Krishna}

\institute
{National Centre for Radio Astrophysics, 
TIFR, Pune University Campus,  Pune 411007, 
India
\and
Department of Physics, University of Pune, Pune 411007, India
\and
Dept.\ of Physics and Astronomy, MSC 8R0314, Georgia State University, Atlanta,
       GA, 30303-3088, USA
}

\offprints{Gopal-Krishna\\
e-mail: krishna@ncra.tifr.res.in}
%\email{krishna@ncra.tifr.res.in}

\date{Received  23 April 2003 / Accepted 22 August 2003}

\titlerunning{Isophotal twists of radio-loud ellipticals}
\authorrunning{Gopal-Krishna et al.}

%\maketitle
%\markboth{}{}
%\begin{abstract}
\abstract{
Using the surface photometric analysis data for a well defined sample of 
79 nearby radio galaxies ($z \le 0.12$), it is shown that the level of 
radio emission
associated with an elliptical galaxy during the radio-loud phase is 
related to the presence of large amounts of isophotal twist in its
optical image. In particular, radio galaxies with $P_{\rm 408} > 3 \times 10^{25}$
W Hz$^{-1}$ appear to show a preference to be associated with
elliptical hosts exhibiting an isophotal twist in excess of $\sim 20^{\circ}$.

}
%\end{abstract}
\maketitle

%\maketitle

\section{Introduction}

Whereas supermassive black holes are now known to reside in the
nuclei of many elliptical galaxies (e.g.\ Magorrian et al.\ 1998;
Gebhardt et al.\ 2000; Ferrarese \& Merritt 2000), % ; Ferrarese 2002)
only a tiny fraction
of them hosts powerful (twin-lobed) radio sources. What triggers the
nuclear activity and how it is related to the actual formation and
evolution of the host galaxy are widely debated issues. 
It is well established that powerful radio emission is almost
exclusively associated with the spheroidal component of galaxies.
However, at any given epoch, an overwhelming majority of luminous
spheroidal galaxies sampled is found to be radio-quiet (e.g., Wisotzki et al.\ 
 2001). One oft discussed possibility, based on 
direct observations, is that galaxy 
mergers or interactions could be playing an important role in radio
activity, as indicated 
by the frequent presence of optical isophotal distortions, shells and 
dust lanes seen in radio-loud ellipticals, particularly in the more 
powerful Fanaroff-Riley (1974) Class II (FR II) radio galaxies (RGs)
(e.g., Heckman et al.\ 1986). It has also been
proposed that the formation of the weaker FR I RGs may 
involve merging of 
early-type galaxies deficient in dense ISM, as the FR I RGs are often
located in richer environments
(Colina \& P{\'e}rez-Fournon 1990; Colina \& de Juan 1995). 

Another set of potentially important clues 
reported on the basis of morphological analyses of the optical 
images are:
(i) a marked tendency for the radio-loud ellipticals to be rounder
than normal ellipticals (Disney et al.\ 1984; Calvani et al.\ 1989);
 (ii) a strong correlation between {\it boxiness} of the elliptical
host and its radio emission (Bender et al.\ 1987, 1989).
Both of these issues are addressed in a more recent study based on a
surface photometric analysis of a well defined sample of 79 radio 
galaxies at $z \le 0.12$ (Govoni et al.\ 2000, hereafter GFFS).
Their analysis, however, did not confirm the suggestion that radio-loud 
ellipticals are rounder in shape, in agreement with the conclusions also 
reached by Smith \& Heckman (1989) for powerful RGs, by 
Ledlow \& Owen (1995) for (mostly) FR I RGs in Abell clusters, 
 and by Falomo et al.\ (2000) for the hosts of BL Lacertae objects. 
Secondly, GFFS find no clear correlation 
between radio emission and the presence of {\it boxy} isophotes in the 
host, in agreement with Ledlow \& Owen (1995) (see also, Jorgenson et al.\ 1995;
Gonz{\'a}lez-Serrano et al.\ 1993).
Another potentially interesting result emerging from their study concerns
the presence of non-concentric disposition of the isophotes, which is
believed to be a strong measure of galaxy interaction. In this respect,
they find no significant difference between the FR I and FR II members
of their sample of radio galaxies.

Larger isophotal twist is another potential indicator of strong galaxy 
interaction or merger events, although a small amount of isophotal twist
could arise simply from the triaxiality of the galaxy (e.g., Kormendy 1982). 
The twist is quantified in terms of the total position angle variation,
$\Delta PA$, of the major axis of the stellar component over the entire 
range in surface brightness. However, isolated ellipticals are found to 
show a similar distribution of twist angle as a randomly selected sample 
of ellipticals, suggesting that by and large, twisting is probably an intrinsic 
property 
(Fasano \& Bonoli 1989; also, Falomo 2003). 

\section{Is the radio output of radio-loud ellipticals related to isophotal twist?}

Recently, Govoni et al.\ (2000) have reported a two-dimensional surface 
photometry analysis for 79 ellipticals with $z < 0.12$, extracted from two 
complete samples of radio galaxies. Among other parameters, 
they have published for each elliptical the value of its ellipticity, 
$\epsilon = 1 - b/a$, measured at its effective radius. Also, for 78 out of the 79 
ellipticals, they have reported the isophotal twist angle, $\Delta PA$. 
This is the maximum twist observed over the range of isophotes. (Note that 
they have excluded both the isophotes at radii $ < 5^{\prime \prime}$ 
as they are usually affected by seeing related smearing as well as
the values of $PA$s having errors larger than $8^{\circ}$.) 
These superior data support the earlier claim that, statistically, rounder 
radio galaxies tend to have larger isophotal twists (Galletta 1980; Fasano \& Bonoli 1989).  
As seen from the $\epsilon$ versus $\Delta PA$ diagram (Fig.\ 16 of GFFS), 
essentially all the elongated ellipticals ($\epsilon > 0.3$) have 
$\Delta PA < 10^{\circ}$. In contrast, for less elongated ellipticals 
($\epsilon < 0.3$), the amount of isophotal twist appears to be independent 
of ellipticity (although few galaxies have $\Delta PA > 40^{\circ}$). 
The origin of this trend is unclear.

The median value $\Delta PA$ for the GFFS sample is found to be $\sim 10^{\circ}$,
similar to $\Delta PA \sim 16^{\circ}$ found for a sample of isolated ellipticals 
using an identical method (Fasano \& Bonoli 1989). Also, at least for 
$z < 0.2$, there is no significant statistical difference between the 
isophotal twist for radio galaxies and radio-quiet ellipticals (Falomo 
et al.\ 2000). Thus, it would appear that radio emission has 
no obvious dependence on isophotal twisting. In contrast, Colina \& de Juan 
(1995) found a significantly larger twisting for FR I radio galaxies,
with a median twist angle of $\sim 14^{\circ}$, compared to a sample of isolated 
normal ellipticals studied by Sparks et al.\ (1991) for which the
median twist angle is only $\sim 4^{\circ}$. 

To follow up these rather conflicting 
claims, we have examined the distribution of $\Delta PA$ against $P_{408}$ for 
the GFFS sample (Fig.\ 1). The values of these two parameters are directly
taken from GFFS, so we follow them in taking $H_0$ = 50 km s$^{-1}$ Mpc$^{-1}$
and $q_0$ = 0. For two of the 78 radio
galaxies, the values of $P_{408}$ are not tabulated by GFFS and we have
determined these by extrapolating from the values of $P_{2700}$ given in GFFS and
$P_{843}$ reported by Jones \& McAdam (1992).

%\vspace{-5.0cm}
\begin{figure}
% THE FOLLOWING COMMAND DOESN'T WORK WELL WITH THIS FILE
\resizebox{\hsize}{!}{\includegraphics{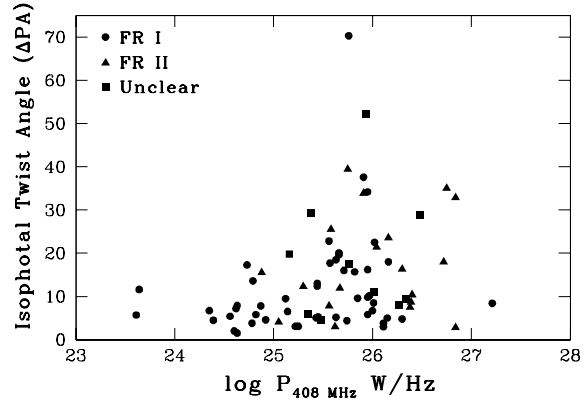}}
%
% THE FOLLOWING THREE COMMANDS ARE APPROPRIATE FOR FINAL 
% TWO COLUMN VERSION
%\vspace{-7.5cm}
%\hspace{-1.5cm}
%\includegraphics[width=10.5cm,height=10.5cm]{fig1.eps}
% THE FOLLOWING THREE COMMANDS ARE APPROPRIATE FOR  
% REFEREE VERSION
%\vspace{-11.5cm}
%\hspace{-1.5cm}
%\includegraphics[width=15.0cm,height=12.0cm]{fig1.eps}
%
\caption{Isophotal twist angles against radio power for this sample.
FR I sources are denoted by circles, FR II's by triangles,
and intermediate or unclear sources by squares.}
\label{fig1}
\end{figure}

%\vspace{-5.0cm} 
\begin{figure}
\resizebox{\hsize}{!}{\includegraphics{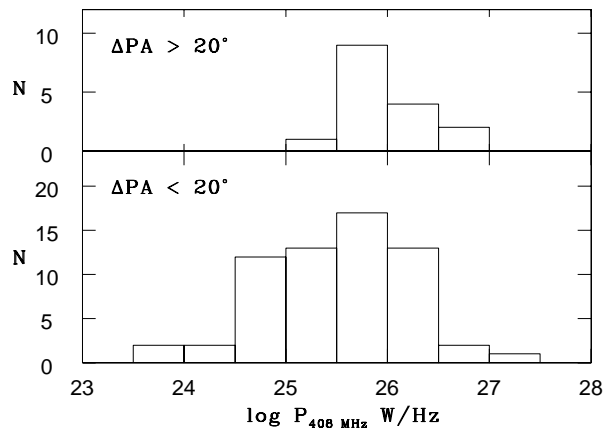}}
%\vspace{-7.0cm}
%\hspace{-1.0cm}
%\includegraphics[width=13.5cm]{fig2.eps}
\caption{Distributions of large ($\Delta PA > 20^{\circ}$) and small ($\Delta PA < 20^{\circ}$)
isophotal twists against radio power.}
\label{fig2}
\end{figure}

From Fig.\ 1, one notices an upper envelope of the data points, 
which rises with the radio luminosity until $P_{408} \sim 10^{26}$ 
W Hz$^{-1}$. 
In particular, larger twist angles ($\Delta PA > 20^{\circ}$) are found almost 
exclusively above $P_{408} \simeq 3 \times 10^{25}$, which is close to the well known 
average transition luminosity between the low and high power radio sources 
(Fanaroff \& Riley 1974). 
We note, however, that  there is a
considerable overlap between the morphologically determined FR I and
FR II sources in the luminosity range considered here (e.g.\
Ledlow \& Owen 1995).  In fact,
as can be seen from Fig.\ 1, there is no obvious relation between
twist angle and FR classification; in order to avoid introducing 
possible bias into the present analysis, we have simply adopted the 
FR classification given by GFFS. Govoni et al.\ (2000) 
themselves note that the
mean twist for FR I's is $13^{\circ} \pm 12^{\circ}$ while for FR II's it
is nearly identical: $15^{\circ} \pm 12^{\circ}$.

Another version of Fig.\ 1 
is displayed in Fig.\ 2 where the histograms of $P_{408}$ are presented for 
two ranges of the twist angle separated at $\Delta PA$ = $20^{\circ}$. 
A Kolmogorov-Smirnov test shows that the two distributions are 
statistically different at the $97\%$ confidence level.
Thus, there seems to be a relation between isophotal twist and
radio luminosity for massive ellipticals in their active phase,
%isophotal twist is a manifestation of the radio luminosity, 
and the two may share a common origin.

Here it may be recalled that in any flux limited sample, such as the
the present one (GFFS), luminosity is strongly correlated with redshift. 
It is therefore conceivable that the primary correlation is between the 
parameters $\Delta PA$ and $z$. However, this is highly unlikely, since that 
would imply an unphysically steep cosmological evolution of $\Delta PA$,
given that the entire sample is defined over a very narrow range in 
redshift ($z \le 0.12$). Explicitly, a plot of twist angle against
redshift shows that sources with $\Delta PA > 20^{\circ}$ are essentially
uniformly distributed in $z$. Alternatively, the radio power could be 
correlated primarily with the rounder shape of the host (Sect.\ 1), 
given the $\Delta PA$ -- $\epsilon$ dependence mentioned above. 
 However, as also 
concluded by Govoni et al.\ (2000), we find no significant correlation 
of radio power
with ellipticity of the host to be evident in the GFFS sample.  Thus, 
it appears more likely that, during the radio-loud phase, 
powerful radio sources are preferentially associated with ellipticals 
exhibiting relatively large isophotal twist. At the same time, a large twist 
cannot be a sufficient condition for generating and sustaining strong 
nuclear activity, since similarly large twists are 
also seen in radio-quiet ellipticals (Falomo et al.\ 2000; Fasano 
\& Bonoli 1989).

In summary, the relation between radio power and the isophotal 
twist reported here suggests that during the radio-loud phase of an
elliptical galaxy the conditions giving rise to a more powerful radio
source are reflected in the occurrence of larger isophotal twist of the 
stellar component. At present, no satisfactory theoretical explanation is 
available for this trend. 
%While it is certainly conceivable 
%that a recent merger or strong tidal interaction could both twist the 
%isophotes and drive more gas into the central engine, given the 
%contradictory evidence for other clear indicators of such a connection,
%this hypothesis currently is not strongly supported. 
It is certainly conceivable 
that a recent merger or strong tidal interaction could both twist the 
isophotes and drive more gas into the central engine, thereby 
triggering the radio activity, but no convincing
results on the mechanisms and relative timescales of these 
phenomena are yet available.
We also note that there is some
support for the scenario in which radio loudness is linked to the
spin of the nuclear black hole (Wilson \& Colbert 1995; also, Small \&
Blandford 1992; Rees 1978).
Thus it is of considerable interest to investigate any possible link between
the isophotal twist of the stellar body and the process of spinning up
the supermassive black hole at the nucleus. It is clearly also important
at this stage to verify the empirical trend emerging in the present 
analysis, using an independent sample of radio galaxies.

\begin{acknowledgements}
We thank Prof.\ Vasant Kulkarni for helpful comments and the anonymous
referee for suggestions that improved the clarity of the manuscript.  PJW
is grateful for hospitality at Princeton University Observatory
and for support from RPE funds at GSU.
\end{acknowledgements}

\end{document}